\documentclass[aps,prl,reprint,showpacs,floatfix,superscriptaddress]{revtex4-1}
\usepackage{graphicx}
\usepackage{amsmath,amsfonts,amscd, physics}
\usepackage{inputenc}
\usepackage{mathrsfs, dsfont}
\usepackage{mathtools}
\usepackage{stmaryrd}
\usepackage{array}
\usepackage{bm}
\usepackage{setspace}
\usepackage[T1]{fontenc}
\usepackage{txfonts}
\usepackage{lineno}
\usepackage{csquotes}

\begin{document}
\title{Quantum Fisher Information Perspective on Sensing in Anti-PT Symmetric Systems}
\author{J. Wang}
\affiliation{Department of Physics and Astronomy, Texas A\&M University, College Station, Texas 77843, USA}
\affiliation{Institute for Quantum Science and Engineering and Department of Biological and Agricultural Engineering, Texas A\&M University, College Station, Texas 77843, USA}
\author{D. Mukhopadhyay}
\affiliation{Department of Physics and Astronomy, Texas A\&M University, College Station, Texas 77843, USA}
\affiliation{Institute for Quantum Science and Engineering and Department of Biological and Agricultural Engineering, Texas A\&M University, College Station, Texas 77843, USA}
\author{G. S. Agarwal}
\affiliation{Department of Physics and Astronomy, Texas A\&M University, College Station, Texas 77843, USA}
\affiliation{Institute for Quantum Science and Engineering and Department of Biological and Agricultural Engineering, Texas A\&M University, College Station, Texas 77843, USA}
\affiliation{Department of Biological and Agricultural Engineering, Texas A\&M University, College Station, Texas 77843, USA}

\begin{abstract}
The efficient sensing of weak environmental perturbations via special degeneracies called exceptional points in non-Hermitian systems has gained enormous traction in the last few decades. However, in contrast to the extensive literature on parity-time (PT) symmetric systems, the exotic hallmarks of anti-PT symmetric systems are only beginning to be realized now. Very recently, a characteristic resonance of vanishing linewidth in anti-PT symmetric systems was shown to exhibit tremendous sensitivity to intrinsic nonlinearities. Given the primacy of sensing in non-Hermitian systems, in general, and the immense topicality of anti-PT symmetry, we investigate the statistical bound to the measurement sensitivity for any arbitrary perturbation in a dissipatively coupled, anti-PT symmetric system. Using the framework of quantum Fisher information and the long-time solution to the full master equation, we analytically compute the Cram\'er-Rao bound for the system properties like the detunings and the couplings. As an illustrative example of this formulation, we inspect and reaffirm the role of a long-lived resonance in dissipatively interacting systems for sensing applications.
\end{abstract}

\maketitle

\section{I. INTRODUCTION}

The Hamiltonian of a physical system characterizes its energy spectrum and time evolution, and is thus, of fundamental importance in quantum theory. In nature, all systems are invariably dissipative as they interface with their environment and relax into thermal equilibrium. As a consequence, an open-system formulation in terms of the master equation becomes imperative. Such a description leads to the system dynamics being described in terms of an effective non-Hermitian Hamiltonian \cite{bender1, bender2}. Over the last few years, certain symmetry properties of these non-Hermitian Hamiltonians have facilitated enhanced sensing of weak perturbations. For instance, the sensing potential of exceptional points (EPs) vis-\`a-vis linear perturbations in both parity-time (PT) symmetric and anti-PT symmetric systems has been brought to the fore \cite{wiersig1, wiersig2, hodaei, demnchrist, lyang, miri, zhang}. For example, in a two-mode system, the square-root singularity at an EP engenders a heightened sensitivity in the spectral splitting to perturbative influences. That is to say, the normal mode splitting scales as the square root of the perturbation, which produces a divergent behavior in the first-order derivative with respect to the perturbation parameter. Recently, in the specific context of anti-PT symmetric systems, a protocol to efficiently detect weak nonlinearities was also proposed \cite{nair}. The scheme exploited a real spectral singularity of the linear response function which strongly inhibits the linewidth of a resonance, thereby drawing out a spectacular response. It was shown that, near the singular point, the nonlinear response $\mathscr{R}$ behaved as $\delta\mathscr{R}/\delta U\propto U^{-5/3}$, where $U$ quantifies the strength of Kerr nonlinearity. Unlike PT symmetry, anti-PT symmetry does not require any extrinsic gain, and is predisposed to simple laboratory realization. A wide array of physical systems have been tailored to exhibit anti-PT symmetry in a multitude of settings including, but not limited to, rapidly moving atomic vapor \cite{peng}, dressed atomic lattices \cite{wu1, wu2}, driven cold atoms \cite{wang, chuang}, laser-cooled atomic ensembles \cite{jiang}, metamaterials \cite{ge}, nonlinear optical systems \cite{anton}, waveguides \cite{konotop,yang}, electrical LRC resonators \cite{choi}, diffusive systems \cite{li}, coupled cavity-magnon systems \cite{hu} and many more, revealing exotic physical phenomena. Recently, Jingwei \textit{et. al.} \cite{wen} demonstrated anti-PT symmetry in a quantum circuit model with three qubits using nuclear spins.

In the context of sensing, it must be acknowledged that the developments heretofore on the sensing potential near special singularities of the system, such as the EP, have been quite remarkable. It is also to be noted that empirical measurements suffer inherently from statistical imprecision, and therefore, uncertainties in the measured response impose fundamental restrictions to the estimation of system parameters. For example, the indelible impact of quantum-limited intensity noise on EP-based sensing protocols was acknowledged in a recent comment by Wiersig \cite{wiersig3}. In order to address the inevitability of statistical errors, we bring in a quantum statistical outlook to the problem of sensing, framed and contextualized in anti-PT symmetric configurations. Employing the well-known framework of quantum Fisher information (QFI) \cite{fisher}, we formulate the sensitivity measure in terms of the Cram\'er-Rao bound \cite{cramer, rao}, which sets down a theoretical constraint to parameter estimation. The concept of Fisher information was conceived as a tool to quantify the amount of information encoded in an observable quantity about an unknown variable through statistical modeling. With the statistical properties of a quantum system described by its density matrix, the notion was subsequently generalized to the quantum formalism \cite{helstrom, holevo, caves2, davi1, davi2, davi3, braun2, jordan, chen, liu}, and has been applied for diverse practical ends, such as for quantum metrology in lossy open systems \cite{paris, braun1, souza, jiaxuan}. In this work, we report the exact precision bound to parameter estimation in anti-PT symmetric systems and thereby, provide an overarching statistical framework for the sensing of weak perturbative effects.

The paper is structured as follows. In section II, we briefly overview the basics of anti-PT symmetry from a classical perspective. In section III, we employ the quantum apparatus and show that anti-PT symmetric systems engineered through vacuum-induced dissipative couplings relax into a coherent state in the long-time limit. In section IV, we work out a general formula for the QFI of a coherent state which provides a lower bound to the sensitivity of an arbitrary perturbation variable.

\section{II. SENSING CAPABILITIES OF CLASSICAL ANTI-PT SYMMETRIC SYSTEMS}

According to the coupled-mode theory in classical optics, the temporal evolution of an open system of two harmonic modes can be described via a $2\cross 2$ non-Hermitian Hamiltonian matrix $\mathscr{H}$. In response to any electromagnetic excitation, the dynamics of the system, in the rotating frame of the drive, can be encoded as
 \begin{equation}
 \begin{split}
 \begin{pmatrix}
\dot{\alpha_0} \\
\dot{\beta_0}
\end{pmatrix} =-i\mathscr{H}\begin{pmatrix}
\alpha_0 \\
\beta_0
\end{pmatrix}+\mathcal{E}\begin{pmatrix}
1 \\
0
\end{pmatrix},
\end{split}
 \end{equation}
where $(\alpha_0, \beta_0)$ signifies the classical amplitudes of the two modes, $\mathcal{E}$ is some generalized Rabi frequency, and $\omega_d$ is the laser frequency. The generic $2\cross 2$ matrix $\mathscr{H}$ lends itself to two very interesting symmetries, namely {PT}-symmetry, which is defined  by the condition $[PT,H]=0$, and anti-{PT} symmetry, which subscribes to $\{PT,H\}=0$. Throughout this manuscript, we tailor our analysis around the constraint of anti-PT symmetry as we set forth some information theoretic restrictions to sensing in systems exhibiting anti-PT symmetry. Such symmetry can be reconciled with the parameter description $\mathscr{H}_{11}=\Delta-i\gamma$, $\mathscr{H}_{22}=-\Delta-i\gamma$, $\mathscr{H}_{12}=\mathscr{H}_{21}=-i\Gamma$, where $(\Delta,\gamma,\Gamma)$ are all real. The coherence-inducing off-diagonal elements of $\mathscr{H}$ characterize a purely dissipative form of interaction between the two modes. The diagonal terms indicate equal damping rates but opposite detunings in the two modes. The eigenvalues for this Hamiltonian are obtained to be $-i\gamma\pm \sqrt{\Delta^2-\Gamma^2}$ for $\abs{\Delta}>\Gamma$ and $-i\gamma\pm i\sqrt{\Gamma^2-\Delta^2}$ (broken anti-PT) for $\abs{\Delta}<\Gamma $. The points of transition $\Delta=\pm\Gamma $, where the eigenvalues coalesce, are known as exceptional points, which have been found useful in the context of sensing. In addition to that, such systems possess a real singularity at $\Delta \rightarrow 0$, $\gamma \rightarrow \Gamma $, which has immense sensing potential, allowing efficient detection of both linear and nonlinear perturbations in the system. The key measurement parameter, in this case, is the residual response of the system characterized by the amplitudes $\alpha_0(\varepsilon)$ and $\beta_0(\varepsilon)$, assumed to be functions of a perturbation $\varepsilon$. Subject to the fulfillment of the stability criterion, \textit{viz.} $\gamma>\Gamma$, the system reduces to a steady state in the long-time limit, yielding the amplitudes
\begin{align}
\alpha_0&=-i\frac{{\mathscr{H}_{22}}}{\det\mathscr{H}}\mathcal{E}=-\frac{\gamma-i\Delta}{\Gamma^2-\Delta^2-\gamma^2}\mathcal{E}, \notag\\
\beta_0&=i\frac{\mathscr{H}_{21}}{\det\mathscr{H}}\mathcal{E}=\frac{\Gamma}{\Gamma^2-\Delta^2-\gamma^2}\mathcal{E}.
\end{align}
In realistic scenarios, however, there would be some discord with the watertight conditions of anti-PT symmetry, no matter how small the error can be made. Since the ensuing mode amplitudes are proportional to $\mathcal{E}$ and sensitive to the system parameters $(\Delta, \gamma, \Gamma)$, the application of a probe field yields nonzero values of the derivatives $\frac{\partial (\alpha_0, \beta_0)}{\partial (\Delta, \gamma, \Gamma)}$. This gives us a way of sensing any small perturbations to these parameters. Let us consider, for instance, the particular case when one of the frequency detunings is zero but there is only a small mismatch in the magnitudes of the two detunings, i.e., $\mathscr{H}_{11}=-i\gamma$ and $\mathscr{H}_{22}=-s-i\gamma$, where $s$ is a small parameter. One could get an estimate of this mismatch by measuring the complex amplitudes 
\begin{align}
\alpha_0(s)&=-\frac{\gamma-is}{\Gamma^2+is\gamma-\gamma^2}\mathcal{E},\notag\\
\beta_0(s)&=\frac{\Gamma}{\Gamma^2+is\gamma-\gamma^2}\mathcal{E}.
\end{align}
The pertinent sensitivities could, then, be obtained in terms of 
\begin{align}
\frac{\partial \alpha_0(s)}{\partial s}&=\frac{i\Gamma^2}{(\Gamma^2+is\gamma-\gamma^2)^2}\mathcal{E},\notag\\
\frac{\partial \beta_0(s)}{\partial s}&=\frac{-i\Gamma\gamma}{(\Gamma^2+is\gamma-\gamma^2)^2}\mathcal{E}.
\end{align}
Clearly, the coherence-induced singularity in the limit $\gamma \rightarrow \Gamma$ gets translated into an infinite theoretical potential for sensing any arbitrary perturbation around this singularity. Supposing that $\gamma=(1+\xi)\Gamma$ for $\xi\ll 1$, both the derivatives would scale as $1/s^{2}$, if $\xi\ll s/\Gamma$. This would imply an augmented sensing capability around this point.
Alternatively, there could be some weak dispersive coupling between $a$ and $b$, which modifies the off-diagonal terms in $\mathscr{H}$ into $\mathscr{H}_{12}=\mathscr{H}_{21}=g-i\Gamma$, where $g\ll\Gamma$. An example of this case from integrated photonics would be a string of two dipolar emitters embedded onto a waveguide and separated by a distance that is scale-wise similar to the resonant wavelength. Following the same approach as above, we can compute the corresponding sensitivities as
\begin{align}
\frac{\partial \alpha_0(g)}{\partial g}&=2i\frac{(\Gamma+ig)(\gamma-i\Delta)}{[(\Gamma+ig)^2-\Delta^2-\gamma^2]^2}\mathcal{E},
\notag\\
\frac{\partial \beta_0(g)}{\partial g}&=-i\frac{(\Gamma+ig)^2+\Delta^2+\gamma^2}{[(\Gamma+ig)^2-\Delta^2-\gamma^2]^2}\mathcal{E}.
\end{align}
Once again, for $\xi\ll g/\Gamma$, the derivatives scale as $1/g^2$ in the limit $\Delta \rightarrow 0$. Both these observations are, however, not only classical but also free from the nuance of statistical considerations. In the subsequent sections, we delve into the quantum statistical framework for treating anti-PT symmetric systems and examine the sensing scheme from the perspective of the QFI. 

\section{III. THE QUANTUM MODEL: STEADY-STATE RESPONSE} The sensitivity bounds, termed as the Cram\'er-Rao bounds, pertaining to precision measurements are furnished in terms of the QFI of the system, thereby necessitating a quantum formulation. In the quantum mechanical picture, the dynamics of a non-interacting two-mode system is underpinned by the Hamiltonian operator
\begin{equation}
\begin{split}
\mathcal{H}/\hbar= \omega_{a}a^{\dagger}a+\omega_{b}b^{\dagger}b,
\end{split}
\end{equation}
where $\omega_{a}$ and $\omega_{b}$ denote the bare resonances of the individual modes $a$ and $b$. If mode $a$ is now irradiated by a laser drive of Rabi frequency $\mathcal{E}$, the interaction Hamiltonian can be reduced to the form
\begin{align}
\mathcal{H}_{int}/\hbar=i(\mathcal{E}e^{-i\omega_dt}a^{\dagger}-\mathcal{E}^*e^{i\omega_dt}a).
\end{align}
To stamp out the harmonic time variation in the interaction, it is convenient to rotate variables as $(a, b)\rightarrow (a, b)e^{-i\omega_dt}$, leading to a new Hamiltonian 
\begin{align}
\mathcal{H}_{eff}/\hbar=\Delta_{a}a^{\dagger}a+\Delta_{b}b^{\dagger}b+i(\mathcal{E}a^{\dagger}-\mathcal{E}^*a).
\end{align}
Now, the choice $\Delta_a=-\Delta_b=\Delta$ would entail a partial fulfillment of the criteria for anti-PT symmetry. In addition to this, both the modes could be made to interface with dissipative reservoirs, which can be appropriately addressed in the master equation formalism. If both the modes decay at identical rates, anti-PT symmetry would be consummated. We assume, for simplicity, thermalized reservoirs at zero temperature instill vacuum noise into the hybrid system. While the noise terms average out to zero, an effective dissipative coupling is introduced between $a$ and $b$. For a system with purely dissipative coupling, the master equation can be expressed as
\begin{equation}
\frac{\dd \rho}{\dd t}=-\frac{i}{\hbar}[\mathcal{H}_{eff},\rho]+\kappa\mathcal{L}(a)\rho+\kappa\mathcal{L}(b)\rho+2\Gamma\mathcal{L}(c)\rho,
\end{equation}
where $\kappa$ stands for the intrinsic damping rate of either of the modes into its local heat bath, $\mathcal{L}$ is the Liouvillian operator defined by $\mathcal{L}(\alpha)\rho=2\alpha\rho \alpha^{\dagger}-\alpha^{\dagger}\alpha\rho-\rho \alpha^{\dagger}\alpha$ for any annihilation operator $\alpha$, and $c=(1/{\sqrt{2}})(a+b)$ is the jump operator corresponding to symmetrical couplings of the modes to a shared reservoir. This equation holds under the approximation that the phase delay due to light propagation from one mode to another is an integral multiple of $2\pi$. We first bring out the correspondence between the quantum mechanical mean values and the classical amplitudes in the steady state. Since $\dot{\expval{a}}=\Tr(\dot{\rho}a)$, the preceding equation naturally dictates the dynamics of the mean values, which, upon simplification, reduces to a form commensurate with Eq. (1):
\begin{equation}
 \begin{split}
 \begin{pmatrix}
\dot{\expval{a}} \\
\dot{\expval{b}}
\end{pmatrix} =-i\mathscr{H}\begin{pmatrix}
\expval{a} \\
\expval{b}
\end{pmatrix}+\mathcal{E}\begin{pmatrix}
1 \\
0
\end{pmatrix},
\end{split}
 \end{equation}
where $\mathscr{H}=\begin{pmatrix}
\Delta-i(\kappa+\Gamma) & -i\Gamma \\
-i\Gamma & -\Delta-i(\kappa+\Gamma) 
\end{pmatrix}$. The identification $\gamma=\kappa+\Gamma$ establishes perfect parity with the classical result. 

Next, we derive the steady state which would accord not only the solutions to the mean values but also the relevant quantum fluctuations. This would be achieved by solving for the full density matrix of the system. Assuming that a steady state exists, this can be worked out by letting $\dot{\rho}=0$. To that end, we first argue that the excitation term $\mathcal{H}_{int}$ gets reflected as a two-mode displacement operation, underscored by the transformation $(a, b)\rightarrow D(\alpha_0,\beta_0) (a,b)D^{\dagger}(\alpha_0,\beta_0)=(\mathcal{A},\mathcal{B)}=(a-\alpha_0,b-\beta_0)$, where $(\alpha_0, \beta_0)=(\expval{a}_{ss}, \expval{b}_{ss})$ and $D(\alpha,\beta)=\exp(\alpha_0 a^{\dagger}-\alpha_0^*a)\exp(\beta_0 b^{\dagger}-\beta_0^* b)$, and $(\alpha_0,\beta_0)$ are given by
\begin{align}
\begin{pmatrix}
\alpha_0\\
\beta_0\\
\end{pmatrix}=
\begin{pmatrix}
\expval{a}_{ss}\\
\expval{b}_{ss}\\
\end{pmatrix}=-i\mathscr{H}^{-1}\begin{pmatrix}
\mathcal{E}\\
0\\
\end{pmatrix}.
\end{align}
This is because, such a transformation dispels the driving term in the master equation, and casts Eq. (9) into the form
\begin{align}
\frac{\dd \rho}{\dd t}=&-\frac{i\Delta}{\hbar}[a^{\dagger}a-b^{\dagger}b,\rho]+\kappa\mathcal{L}(\mathcal{A})\rho+\kappa\mathcal{L}(\mathcal{B})\rho+2\Gamma\mathcal{L}(\mathcal{C})\rho\notag\\
=&-(\gamma+i\Delta)[\mathcal{A}^{\dagger},\mathcal{A}\rho]-(\gamma-i\Delta)[\mathcal{B}^{\dagger},\mathcal{B}\rho]\notag\\
&-\Gamma[\mathcal{A}^{\dagger},\mathcal{B}\rho]-\Gamma[\mathcal{B}^{\dagger},\mathcal{A}\rho] +\textit{h.c.},
\end{align}
with $\textit{h.c.}$ denoting the Hermitian conjugate, and $\mathcal{C}=(1/{\sqrt{2}})(\mathcal{A}+\mathcal{B})$. Now, in view of the uniqueness of the steady state, we obtain a solution embodied by the conditions
\begin{align}
\mathcal{A}\rho=\rho \mathcal{A}^{\dagger}=0,\notag\\
\mathcal{B}\rho=\rho\mathcal{B}^{\dagger}=0,
\end{align}
which are conformable only with the bimodal vacuum $\rho=\ket{0_{\mathcal{A}},0_{\mathcal{B}}}\bra{0_{\mathcal{A}},0_{\mathcal{B}}}$ in the displaced representation. Reverting to the original representation, we obtain a coherent state
\begin{align}
\rho=\ket{\alpha_0,\beta_0}\bra{\alpha_0,\beta_0}
\end{align}
as our solution to the anti-PT symmetric master equation. In view of Eq. (13), it follows that a calculation of the mean values $\expval{a}$ and $\expval{b}$ with respect to this coherent state coincides with the classical result laid out in Eq. (2).

The above state can also be inferred from the reformulated master equation expressed in terms of the corresponding phase-space distribution $P(\alpha,\beta)$ for a two-mode system. Following the procedure spelled out in \cite{gsabook}, we deduce an equivalent master equation in the $P$-representation as

\begin{align}
\frac{\partial P}{\partial t}&=\frac{\partial}{\partial\alpha}[u(\alpha,\beta)P]+\frac{\partial}{\partial\beta}[v(\alpha,\beta)P]+c.c.\,,
\end{align}

\noindent where $u(\alpha,\beta)=(\gamma+i\Delta)\alpha+\Gamma\beta-\mathcal{E}$, $v(\alpha,\beta)=\Gamma\alpha+(\gamma-i\Delta)\beta$, and  $P(\alpha,\beta)$ is defined as  

\begin{equation}
\rho=\int\int P(\alpha,\beta)\ket{\alpha,\beta}\bra{\alpha,\beta}d^{2}\alpha d^{2}\beta\,,  \label{p}
\end{equation}

\noindent and $c.c.$ stands for the complex conjugate. In the steady state, as $t\rightarrow\infty$, we require $\frac{\partial P}{\partial t}=0$. This provides the quantum steady-state solution for the $P$ function to be

\begin{equation}
P(\alpha,\beta)=\mathscr{N}\delta^{(2)}(u(\alpha,\beta))\delta^{(2)}(v(\alpha,\beta))\,,  \label{psolution}
\end{equation}

\noindent where $\delta(.)$ represents the Dirac delta function and $\mathscr{N}$ is some normalization constant, since $P$ must satisfy $\int\int P(\alpha,\beta)d^{2}\alpha d^{2}\beta=1$. Considering the linearity of the functions $u$ and $v$, this is equivalent to

\begin{equation}
P(\alpha,\beta)=\delta^{(2)}(\alpha-\alpha_{0})\delta^{(2)}(\beta-\beta_{0})\,,  \label{psolution2}
\end{equation}

\noindent where $\alpha_{0}=-\frac{(\gamma-i\Delta)}{\Gamma^2-\Delta^2-\gamma^2}\mathcal{E}$ and $\beta_{0}=\frac{\Gamma}{\Gamma^2-\Delta^2-\gamma^2}\mathcal{E}$ is the solution to $u(\alpha_0,\beta_0)=v(\alpha_0,\beta_0)=0$. Substituting this into Eq. (15), we find that the quantum steady state of this driven anti-PT symmetric system must be a coherent state $\ket{\alpha_{0},\beta_{0}}\bra{\alpha_0,\beta_0}$. The coherent nature of this state would remain valid even for small perturbations to any of the system parameters, as long as the system remains purely dissipative. However, the amplitudes $(\alpha_0, \beta_0)$ would get modified as per the constraint of Eq. (12). In particular, if the two modes address each other through a weak dispersive interaction $\hbar g(a^{\dagger}b+ab^{\dagger})$, the solutions to the response can be found by assigning an imaginary offset to $\Gamma$, i.e., by letting $\Gamma\rightarrow\Gamma+ig$.

\section{IV. THE QUANTUM CRAM\'ER-RAO BOUND FOR SENSITIVITY}
 
As proved in section III, the steady state for an anti-PT symmetric quantum system is a two-mode coherent state that can be written as $\ket{\alpha_{0},\beta_{0}}\bra{\alpha_0,\beta_0}$, where $\alpha_0=-\frac{(\gamma-i\Delta)}{\Gamma^2-\Delta^2-\gamma^2}\mathcal{E}$ and $\beta_0=\frac{\Gamma}{\Gamma^2-\Delta^2-\gamma^2}\mathcal{E}$ are the long-time responses in the two modes induced by the external field. Thus, information about any arbitrary perturbations to the system parameters would be encoded in the complex amplitudes $\alpha$ and $\beta$, and as such, perturbative effects can be sensed via measurements of the response functions. To put this into information-theoretic perspective, we inspect the Cram\'er-Rao bound for the measurement uncertainty in the estimation of a perturbation parameter. The problem can, in fact, be solved in a general way, and applied to any unknown parameter $\varepsilon$, as long as the response functions $\alpha_0(\varepsilon)$ and $\beta_0(\varepsilon)$ are exactly known.

To start off, we briefly recapitulate the basic concepts of QFI. In the estimation of an unknown parameter $\varepsilon$, the  Cram\'er-Rao bound reads $\delta \varepsilon \geq \sqrt{F_{Q}^{-1}(\varepsilon)}$, where $\delta \varepsilon$ symbolizes the measurement uncertainty or sensitivity of $\varepsilon$, and $F_{Q}(\varepsilon)=\Tr[\rho L^{2}]$ is the QFI for $\varepsilon$, pertaining to a system with density matrix $\rho$. The symmetric logarithmic derivative (SLD) matrix $L$ is calculated from the equation $\frac{\partial \rho }{\partial
\varepsilon}=\frac{1}{2}(L\rho +\rho L)$. Since we have proved that the anti-PT symmetric system goes over into a coherent state in the long-time limit, we work out the algebra for the appropriate steady state. The corresponding density matrix can be expanded out in the basis of Fock states as 
\begin{equation}
\rho=e^{-|\alpha_0|^{2}}e^{-|\beta_0|^{2}}\sum_{mnpq}\frac{\alpha_0^{m}\beta_0^{n}\alpha_0^{*p}\beta_0^{*q}}{\sqrt{m!n!p!q!}}\ket{m,n}\bra{p,q}\,.  \label{density}
\end{equation}

\noindent Since each of the parameters $(\alpha_0$, $\beta_0)$ is a function of $\varepsilon$, we can solve directly for the derivative $\frac{\partial\rho_{{mn,pq}}}{\partial\varepsilon}$ in terms of the $\frac{\partial\alpha_0}{\partial\varepsilon}$ and $\frac{\partial\beta_0}{\partial\varepsilon}$. Note that $\alpha_0$, $\beta_0$ are complex numbers. Therefore, expressing them as $\alpha_0=\abs{\alpha_0}e^{i\theta_1}$, $\beta_0=\abs{\beta_0}e^{i\theta_2}$, we find the relation

\begin{align}
\frac{\partial\rho_{{mn,pq}}}{\partial\varepsilon}&=\bigg[m\bigg(\frac{\partial ln|\alpha_0|}{\partial\varepsilon}+i\frac{\partial\theta_{1}}{\partial\varepsilon}\bigg)+n\bigg(\frac{\partial ln|\beta_0|}{\partial\varepsilon}+i\frac{\partial\theta_{2}}{\partial\varepsilon}\bigg)\notag \\  \label{derivative} 
&+p\bigg(\frac{\partial ln|\alpha_0|}{\partial\varepsilon}-i\frac{\partial\theta_{1}}{\partial\varepsilon}\bigg)+q\bigg(\frac{\partial ln|\beta_0|}{\partial\varepsilon}-i\frac{\partial\theta_{2}}{\partial\varepsilon}\bigg)\notag \\ 
&-2|\alpha_0|\frac{\partial|\alpha_0|}{\partial\varepsilon}-2|\beta_0|\frac{\partial|\beta_0|}{\partial\varepsilon}\bigg]\rho_{mn,pq}\,,
\end{align}

\noindent which can be analyzed by resolving them into three parts, defined below:

\begin{align}
&-2\bigg(|\alpha_0|\frac{\partial|\alpha_0|}{\partial\varepsilon}+|\beta_0|\frac{\partial|\beta_0|}{\partial\varepsilon}\bigg)\rho_{mn,pq}=\frac{1}{2}(L_{1}\rho+\rho L_{1})_{mn,pq},\notag\\
&\bigg[(m+p)\frac{\partial ln|\alpha_0|}{\partial\varepsilon}+(n+q)\frac{\partial ln|\beta_0|}{\partial\varepsilon}\bigg]\rho_{mn,pq}=\frac{1}{2}(L_{2}\rho+\rho L_{2})_{mn,pq},\notag\\
&i\bigg[(m-p)\frac{\partial\theta_{1}}{\partial\varepsilon}+(n-q)\frac{\partial\theta_{2}}{\partial\varepsilon}\bigg]\rho_{mn,pq}=\frac{1}{2}(L_{3}\rho+\rho L_{3})_{mn,pq}.  \label{l1}
\end{align}

\noindent Following the method discussed in \cite{jxn2}, we solve the equations for the individual components separately. In other words, the corresponding SLD gets split as 
\begin{equation}
L=L_{1}+L_{2}+L_{3}\,,  \label{l}
\end{equation}
\noindent where

\begin{align}
L_{1}&=-2|\alpha_0|\frac{\partial|\alpha_0|}{\partial\varepsilon}-2|\beta_0|\frac{\partial|\beta_0|}{\partial\varepsilon},\notag\\
L_{2}&=2\frac{\partial ln|\alpha_0|}{\partial\varepsilon}(a^{+}a)+2\frac{\partial ln|\beta_0|}{\partial\varepsilon}(b^{+}b),\notag\\
L_{3}&=2i\frac{\partial\theta_{1}}{\partial\varepsilon}[a^{+}a,\rho]+2i\frac{\partial\theta_{2}}{\partial\varepsilon}[b^{+}b,\rho].
\end{align}

\noindent This precise form of $L$ pins down the QFI and the corresponding sensitivity limit. Invoking the definition $F_{Q}(\varepsilon)=\Tr[\rho L^{2}]$ we conclude that
\begin{equation}
F_{Q}(\varepsilon)=4\abs{\frac{\partial\alpha_0}{\partial\varepsilon}}^{2}+4\abs{\frac{\partial\beta_0}{\partial\varepsilon}}^{2}\,. \label{qfi}
\end{equation}%

The above expression is absolutely general, regardless of the specifics of the response function $(\alpha_0(\varepsilon),\beta_0(\varepsilon))$. Thus, the QFI, as well as the sensitivity for the system parameters, both depend on the derivatives $\abs{\frac{\partial\alpha_0}{\partial\varepsilon}}$ and $\abs{\frac{\partial\beta_0}{\partial\varepsilon}}$. This result is specific to an anti-PT symmetric mode hybridization, when the interposing reservoir has relaxed into the vacuum state at zero temperature. Apropos of our problem, where the coherent state $\ket{\alpha_{0},\beta_{0}}\bra{\alpha_0,\beta_0}$ encapsulates the statistical properties of the measured quantities, we can model the detuning mismatch $s$ or the weak coupling $g$ as a perturbation. Figs 1 (a) and (b) depict the sensitivity bounds for these parameters for a set of varying decay rates. For progressively smaller values of the intrinsic damping rate, the perturbations can be sensed with incrementally larger efficacy near the singular point where both the detunings approach zero.

\begin{figure}[h!]
\centering\includegraphics[width=8cm]{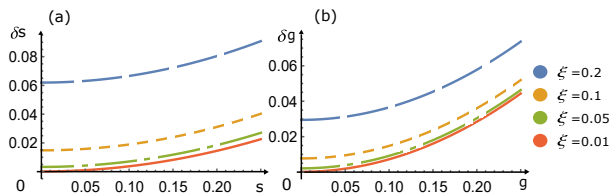}
\caption {The Cram\'er-Rao bounds for measurement uncertainties as $\gamma \rightarrow \Gamma$, scaled in units of $\Gamma$: (a) $\delta s$ for the sensing of detuning mismatch $s$ for different values of $\xi$; (b) $\delta g$ for the sensing of the dispersive coupling strength $g$, at $\Delta = 0.1\Gamma$. The quantity $\xi$, defined as $\xi=(\gamma-\Gamma)/\Gamma$, is a dimensionless measure of the rate of extraneous damping.} 
\label{Fig1}
\end{figure}

\section{VIII. CONCLUSIONS}

In summary, we have offered an information-theoretic insight into the subject of sensing in anti-PT symmetric systems where the coupling between the participating modes is produced by a common vacuum. We invoked the Fisher information theory as a statistical tool to circumscribe the theoretical precision of a single-parameter measurement in terms of the Cram\'er-Rao bound, which serves as a well-rounded metric for the sensitivity. We argued that anti-PT symmetric systems decay into a coherent state in the long-time limit, which renders the QFI calculation tractable. Then, by treating the steady-state response functions as sensing signals, we derived an exact statistical expression for the sensitivity of the perturbation variable. To exemplify the role of the QFI, we applied this approach to the sensing of a detuning mismatch as well as that of weak coherent coupling between the two modes around a real singularity of the hybrid system. The vanishingly small sensitivity bounds enabled by the emergence of a long-lived resonance reinforced the impact of a strong vacuum induced coherence on precision measurements.

\section{ACKNOWLEDGMENTS}

We thank the support of Air Force Office of Scientific Research (Award N%
\textsuperscript{\underline{o}} FA-9550-20-1-0366) and the Robert A Welch
Foundation (A-1943-20210327).

\end{document}